\documentstyle[11pt,moriond,epsfig]{article}

\def\Journal#1#2#3#4{{#1} {\bf #2}, #3 (#4)}

\def\PRD{{\em Phys. Rev.} D}
\def\ZPC{{\em Z. Phys.} C}

\def\be{\begin{equation}}
\def\ee{\end{equation}}
\def\bea{\begin{eqnarray}}
\def\eea{\end{eqnarray}}

\begin{document}
\begin{flushright}
Fermilab-Conf-98/106-T
\end{flushright}
\vspace*{3cm} \title{THEORETICAL UNCERTAINTIES ASSOCIATED WITH THE
EXTRACTION OF $M_W$ AT HADRON COLLIDERS\footnote{Talk presented at the
XXXIIIrd Rencontres de Moriond on "QCD and High Energy Hadronic
Interactions".}}

\author{S.~KELLER}

\address{Theoretical Physics Dept, Fermilab, PO Box 500,\\
Batavia, IL, USA}

\maketitle
\abstracts{In this contribution I briefly summarize several topics
related to the measurement of the W-boson mass, $M_W$, at hadron colliders.}

\section{Introduction}

A precise measurement of $M_W$, along with other measurements like the
mass of the top quark, will indirectly constrain the mass of the
elusive Higgs~\footnote{Currently the measurement of the weak mixing
angle gives a better constraint, see Ref.~\cite{Altarelli}.}, the
missing piece of the very successful Standard Model.  This is
important, an indirect measurement tells us where to look for the Higgs
in direct measurement and later when (if) the Higgs is discovered a
comparison of the direct and indirect measurements will provide an
important test of the Standard Model.  Both the Tevatron and LEP have
already made very precise measurements of $M_W$, as reported in these
proceedings~\cite{current}.  In Table~\ref{tab:cdf2}, I summarize the
(CDF) expectations for the uncertainties on $M_W$ for Run II at the
Tevatron~\cite{cdf2}.
\begin{table}[h]
\caption{Run II expectations for the uncertainties on $M_W$, in MeV,
from Ref.$^3$, for an integrated luminosity of 2 $fb^{-1}$  }
\label{tab:cdf2}
\vspace{0.4cm}
\begin{center}
\begin{tabular}{|c|c|c|}\hline 
Sources of Uncertainties & $W \rightarrow e \nu$ & $W \rightarrow \mu
\nu$ \\ \hline
Statistical & 14 & 20 \\ \hline
W Production Model & 30 & 30 \\ \hline
Other Systematic Uncertainties & 25 & 22 \\ \hline Total
Uncertainty &42 &40 \\ \hline
\end{tabular}
\end{center}
\end{table}
As can be seen in this table, the W production model uncertainty
dominates.  This is the uncertainty associated with the transverse
momentum of the W, the parton distribution functions (PDF's), and the QCD and
electroweak corrections.  The fact that this uncertainty dominates
represents both a challenge and an opportunity.  It is a challenge
because it is not acceptable and we should find ways to decrease this
uncertainty below the experimental uncertainty.  It is an opportunity
because if we successfully decrease it then the Run II measurement of
$M_W$ at the Tevatron will be even better than currently anticipate.
Note also that the W production model uncertainty is fully correlated
between the electron and muon channels, such that not much improvement
is gained by combining the two.  If we succeed in controlling the W
production model uncertainty we could get four (two per detectors)
measurements with uncertainty smaller than 40 MeV, and an overall
uncertainty of the order of 20 MeV might be possible.

In the remainder of this contribution, I review the current status of
the electroweak corrections to Z and W production at hadron colliders,
a ratio method to measure $M_W$ (and $\Gamma_W$), recent developments
on PDF uncertainties, and the opportunity to very precisely measure
$M_W$ at the LHC.  I give my conclusions in the last section.

\section{Electroweak Corrections to Z and W Production}

This section is a summary of the work done in collaboration with
U.~Baur and W.~Sakumoto in Ref.~\cite{Zcalc} (corrections to Z production)
and with U.~Baur and D.~Wackeroth in Ref.~\cite{Wcalc} (corrections to
W production).
There is a shift in $M_Z$ and $M_W$ extracted from the data due to the
electroweak corrections of the order of 100 MeV.  We need to
understand the uncertainty associated with that shift.  The uncertainty was
assumed to be of the order of 20 MeV for RunIa analysis at the Tevatron.

The electroweak corrections to Z production are also needed because the 
measured $M_Z$ and $\Gamma_z$ are used to calibrate the detector when 
compared to the values measured at LEP.

In the calculation used so far to extract $M_W$ (Berends and
Kleiss~\cite{BK}, 1985), only the final state photonic corrections are
included using a very good approximation.  The accuracy of
this approximation can only be estimated by doing the full
calculation.  Our calculations include initial and final state
corrections and their interference.  We used the phase space slicing
method, as in QCD~\cite{NLOMC}; the advantage of that method is that
the experimental cuts can be imposed without any difficulties, without
having to redo analytical integrations.  In the calculations, we kept
the mass of the final state charged lepton(s), it protects the final
state collinear singularities.  The final state photonic corrections
dominate the electroweak corrections because they are enhanced by
$\alpha \: log(M_{Z or W}^2/m_{lepton}^2)$ when the charged lepton and
photon momentum are not recombined.  These large contributions are not
present in the integrated cross section as required by the KLN
theorem~\cite{KLN}.  The universal initial state collinear
singularities have to be absorbed into the PDF's by factorization, in
complete analogy with QCD.  In principle, for the overall consistency
of the calculations, the QED corrections should be added to the
evolution of the PDF's and incorporated into the global fitting of
PDF's.  Because this has not yet been done, we only have partial
information about the impact of the initial state corrections.

In the Z case, the QED corrections are gauge invariant by themselves,
and so far we neglected the weak corrections, they are expected to be
small.  In the W case the QED corrections are not gauge invariant by
themselves, the weak corrections must be included.  The non trivial
calculation of the matrix elements for the W case was done in
Ref.~\cite{DH} by D.~Wackeroth and W.~Hollick.

Our results are showing that, as expected, the final state corrections
dominate the shape change of the distributions in the region of
interest for the measurement of $M_W$.  The most important detector
effect is the recombination: when the electron and the photon are
close to each other then their momenta is recombined to an effective
electron momenta.  This effect reduces the size of the
corrections, although not to a level where they can be neglected.

The most important result is that the Z and W masses obtained by fitting with
our $O(\alpha)$ calculations are about 10 MeV smaller than that
obtained by fitting with the approximate calculations used so far.
This is a good because this 10 MeV shift is smaller than the
uncertainty so far assumed in the analysis.  It is important to
understand that this 10 MeV is NOT the uncertainty on the $O(\alpha)$
calculation, it is simply the difference between two calculations of
the $O(\alpha)$ corrections.  The uncertainty on the $O(\alpha)$
calculation can only be estimated from the size of the $O(\alpha^2)$
corrections.  Now that we have shown that the approximation \`a la
Berends and Kleiss is very good for the $O(\alpha)$ corrections, the
same type of approximation could be used to obtain an estimate of the
$O(\alpha^2)$ corrections.

\section{Ratio Method to Measure $M_W$ (and $\Gamma_W$)}
\label{sec:ratio}

This section is a summary of the work done in collaboration with
W.~Giele in Ref.~\cite{GK1}.  Instead of using the W distribution to
measure $M_W$ and the W width, $\Gamma_W$, the ratio of W over Z
distributions can be used.  The normalization of the ratio should be
included in the fit as it is sensitive to $\Gamma_W$.  This idea is not
really new, after all the measurement of $M_Z$ and $\Gamma_Z$
is already used for calibration of the detectors.  However, with the
upcoming high luminosity run at the Tevatron, the idea can be brought
to full maturity.  The main difference between the W and Z production
is due to their different mass ($M_V$).  Mass-scaled variables must
therefore be used:
\vskip .5cm 
\bea
x=\frac{\cal O}{M_V} \nonumber \\
R= \frac{\left. \frac{d\sigma}{dx}\right|_W}
{\left. \frac{d\sigma}{dx}\right|_Z},
\eea
\vskip .5cm \noindent where ${\cal O}$ is the observable under study.
$M_W$ can be fitted for such that the measured ratio R is equal to the
calculated one.

The obvious limitation of the method is that it depends on the Z 
statistics.  It is about 10 times lower than in the W case, such that the
statistical uncertainty is about $\sqrt{10}$ times larger~\footnote{$\sqrt{10}$
is replaced by $\sqrt{5}$ for observables that depend on one charged lepton, 
such that both leptons in the Z case can be entered in the distribution.}.

There are many advantages to the method.  First, the experimental
systematic uncertainties tend to cancel in the ratio.  Potential
problems that will spoil the cancellation are, e.g., the isolation criteria of 
the 2nd lepton in the Z case and some of
the backgrounds that are different.  Second, $M_W$ and $\Gamma_W$ are
directly measured with respect to $M_Z$ and $\Gamma_Z$ which were
accurately measured at LEP.  Third, the QCD corrections to the ratio
are smaller than for the W and Z observables themselves which means
that the theoretical uncertainty on the ratio is also smaller (we have
checked this statement for the transverse mass, the transverse energy
of the vector boson and the transverse energy of the lepton distributions, see
Ref.~\cite{GK1}).  Finally, the expectation is that the PDF uncertainties
will also be smaller.

In this ratio method, there is a clear trade-off between statistical
and systematic uncertainties: the statistical uncertainty is increased
while the systematic uncertainty is decreased.  We therefore expect
this method to be very competitive at high luminosity 
(Run II, TeV33~\cite{TEV33})
because there the standard method uncertainty is dominated by the
systematic uncertainty, see Table~\ref{tab:cdf2}.

D0 has already applied the ratio method to the transverse mass
distribution with very encouraging results, see Ref.~\cite{D0RATIO}.
The ratio method applied to the transverse energy of the charged
lepton might yield the smallest uncertainty on $M_W$ at high luminosity.

\section{Parton Distribution Function Uncertainties}

This section is a summary of the work done in collaboration with
W.~Giele in Ref.~\cite{GK2}.
Standard sets of PDF's do not come with uncertainties.
The spread between different sets is often associated with PDF
uncertainties.  This is the case for the $M_W$ analysis at the
Tevatron.  As is well known, it is not clear at all what this spread
represents.  It is time for a set of PDF's with uncertainties.
In Ref.~\cite{GK2} we developed a method, within the framework of
statistical inference, to take care of the PDF uncertainties.
Here I simply explain two important steps of the method.

The first step is the propagation of the uncertainty to new observables.
The PDF's are assumed to be parametrized at a scale $Q_0$, with $N$ parameters,
$\{\lambda\}\equiv\lambda_1, \lambda_2, \ldots, \lambda_{N}$.  The
probability density distribution of these parameters, 
$P_{init}(\lambda)$, is also assumed to be known.

For any observable, $0(\lambda)$, the prediction is simply given by the 
average value over the multi-dimensional parameter space:
\vskip .5cm 
\be
<O> = \int_V O(\lambda) P_{init}(\lambda) d\lambda.
\ee
\vskip .5cm To calculate the integral we use a Monte-Carlo approach
with importance sampling.  We generate 100 random sets of parameters
distributed according to the initial probability density distribution,
$P_{init}(\lambda)$.  This corresponds to 100 sets of PDF's that
represent the uncertainty.  The observable can be calculated for each
set, $O^j$, and the prediction is then given by the average value over
the 100 PDF sets:
\vskip .5cm 
\be <O> \approx\frac{1}{100}\sum_{j=1}^{100} O^j, 
\ee
\vskip .5cm \noindent whereas the PDF uncertainty is given by the standard 
deviation, $\sigma_O$, of the 100 PDF sets:
\vskip .5cm
\be
\sigma_O^2 \approx\frac{1}{100}\sum_{j=1}^{100} 
\left(O^j-<O>^2\right)^2 
\ee
\vskip .5cm 
This gives a simple way to propagate the uncertainties to
new observables, in particular there is no need for the derivative of
the observable with respect to the parameters.

The second step I want to describe is the inclusion of the effect of
new data on the PDF's.  If the new data agrees with the prediction
then the effect of the new data can be included by updating the
probability density distribution with Bayes theorem.  Initially, each
of the 100 sets of PDF's ($PDF_i$) has a constant weight because of
the use of importance sampling.  Now each of the sets acquires a
different weight given by the conditional probability density
distribution of the set considering the new data:
\vskip .5cm
\be
P_{new}(PDF_i)=P(PDF_i/ new \, \,  data)
\ee
\vskip .5cm \noindent The latter is directly given by Bayes theorem:
\vskip .5cm 
\be
P(PDF_i/ new \, \,  data) \propto P(new \, \, data/PDF_i)\,\, 
P_{init}(PDF_i)
\ee
\vskip .5cm 
If the uncertainties on the data are Gaussian
distributed, then the weights are given by:
\vskip .5cm 
\be
P(new \, \, data/PDF_i) \propto e^{-\frac{\chi^2_i}{2}}
\ee
\vskip .5cm \noindent where $\chi^2_i$ is the chi-squared of the new
data with the theory calculated with the specific set of
PDF's.  Prediction for yet other observables that includes the effect
of the new data can now be calculated by using weighted sum.  No
information about the data used to derive $P_{init}(\lambda)$ is
needed.  Other advantages of the method are as follows.  The
probability density distribution of the parameters does not have to be
Gaussian.  A data set can be easily excluded from the fit and
experimenters can include their own data into the PDF's during the
analysis phase.  Finally, the theory uncertainty can be easily
included.

It is worth mentioning that S.~Alekhin about a year ago extracted
PDF's with uncertainties from deep inelastic scattering (DIS)
data~\cite{ALEK}.  Both the statistical and systematic uncertainties
with correlations were included.  However the theoretical uncertainty
was not considered.  In Ref.~\cite{GK2} we used his results for our
initial probability density distribution to predict two observables at
the Tevatron: the single inclusive jet cross section and the lepton
charge asymmetry in W decays.  Note that the initial probability
density distribution could also be entirely based on theoretical
consideration, in the spirit of Bayes theorem.  One remaining problem
is the uncertainty associated with the choice of parametrization of
the input PDF's.  This is a difficult problem that does not have a
clear answer and will require a compromise between the number of
parameters and the smoothness of the PDF.

\section{Measurement of $M_W$ at the LHC}

This section is a summary of the work done in collaboration with
J.~Womersley in Ref.~\cite{WK}.
The LHC will be a copious source of W.  The cross section for W production 
(with appropriate cuts) at the LHC is about four times larger than at the 
Tevatron.  The statistical uncertainty should therefore be small.

A priori, the systematic uncertainty is expected to be large at the
LHC.  However it is likely that the LHC will run at ``low'' luminosity
($\sim 10^{33}cm^{-2} s^{-1}$) for at least a year, corresponding to
an integrated luminosity of ${\cal L}=10fb^{-1}$.  At that luminosity
the detector capabilities are very good: triggering on leptons with
transverse energy as low as $\sim 20$ GeV is possible, the number of
interactions per crossing is of the order of 2, providing a quiet
environment, and the missing transverse momentum will be well
measured because the hadronic calorimeters have large coverage (up to
pseudorapidity of 5).  Furthermore, both the ATLAS and CMS detectors
offer advances over their counterparts at the Tevatron for lepton
identification and measurement.

The QCD corrections to the shape of the transverse mass distribution
are of the order of 10\% in the region of interest.  The 
corrections are larger than at the Tevatron ($\sim 2 \%$) but still
reasonable.  The NNLO calculation will be useful in this case.  If
necessary the ratio method, explained in section~\ref{sec:ratio},
could be used to reduce the theoretical uncertainty.

Scaling from the current measurement at the Tevatron, about 
$15 \: 10^6$ $W \rightarrow e \nu$ reconstructed events 
are expected for 10 $fb^{-1}$ at the LHC.
The uncertainty obtained by using the parametrization developed for
the Tev2000~\cite{tev2000} study is very small, of the order of 8 MeV.  
It is difficult to believe that such a small uncertainty will be
reached.  However, we take this as an indication that there is an
opportunity to make the world's best measurement of $M_W$, i.e. to
measure $M_W$ to a precision better than 15 MeV, the goal
of TeV33.

Note also that the Bjorken-x probed is different at the LHC and the 
Tevatron.  Therefore the PDF uncertainty will be different and from that 
point of view the two measurements will be complementary.

\section{Conclusions: Things to do!} 

A precise measurement of $M_W$ will be important to further constrain
the mass of the Higgs.  Current extrapolations to higher luminosity at
the Tevatron indicate that the uncertainty on the extraction of $M_W$
will be dominated by theoretical uncertainties.  We therefore have
work to do to ensure that this does not remain the case.  For example,
the two loop corrections ($O(\alpha_s^2)$, $O(\alpha_s \alpha)$, and
$O(\alpha^2)$) are needed to evaluate the theoretical uncertainty on
the one loop calculations.  A more definite statement about the impact
of the initial state contribution of the electroweak corrections is
needed.  We only have indications that they have a small effect.  DIS
and Tevatron data should be used to extract PDF's with uncertainties
with the method described in Ref.~\cite{GK2}.  A lot of work remains
to be done but the theoretical uncertainty should be significantly
decreased by the time the Tevatron takes data again.

\end{document}